\documentclass[10pt,twocolumn,aps,pra,superscriptaddress,showpacs,tightenlines,pdflatex,longbibliography]{revtex4-2}
\usepackage{upgreek}
\usepackage{amsthm}
\usepackage{gensymb}
\usepackage{braket}
\usepackage{amsmath}            
\usepackage{amssymb}            
\usepackage{mathtools}
\usepackage{graphicx}           
\usepackage[caption=false]{subfig}
\usepackage{xcolor}

\usepackage{etoolbox}
\usepackage{breqn}

\usepackage{enumitem}
\usepackage{xcolor}
\usepackage{bm}

\makeatletter
\let\cat@comma@active\@empty

\usepackage[colorlinks,citecolor=blue,urlcolor=blue,bookmarks=false,hypertexnames=true]{hyperref} 

\newcommand{\proj}[1]{\ket{#1}\bra{#1}}
\newcommand{\ts}{\otimes}
\newcommand{\id}{\openone}
\newcommand{\Tr}[1]{\text{Tr}\left[#1\right]}
\newcommand{\A}{\text{A}}
\newcommand{\B}{\text{B}}
\newcommand{\Q}{\text{Q}}
\newcommand{\C}{\text{C}}

\newcommand{\opt}{\text{opt}}

\newcommand{\red}[1]{{\color{red} #1}}

\newcommand{\comment}[1]{}

\newcommand{\fig}[1]{Fig.~#1}

\begin{document}
\title{Steering-enhanced quantum metrology using superpositions of noisy phase shifts}

\author{Kuan-Yi Lee}
\email{These authors contributed equally.}
\affiliation{Department of Physics and Center for Quantum Frontiers of Research \&
Technology (QFort), National Cheng Kung University, Tainan 701, Taiwan}
\author{Jhen-Dong Lin}
\email{These authors contributed equally.}
\affiliation{Department of Physics and Center for Quantum Frontiers of Research \&
Technology (QFort), National Cheng Kung University, Tainan 701, Taiwan}

\author{Adam Miranowicz}
\affiliation{Theoretical Quantum Physics Laboratory, Cluster for Pioneering Research, RIKEN, Wakoshi, Saitama 351-0198, Japan}
\affiliation{Institute of Spintronics and Quantum Information, Faculty of Physics, Adam Mickiewicz University, 61-614 Poznań, Poland}

\author{Franco Nori}
\affiliation{Theoretical Quantum Physics Laboratory, Cluster for Pioneering Research, RIKEN, Wakoshi, Saitama 351-0198, Japan}
\affiliation{Center for Quantum Computing, RIKEN, Wakoshi, Saitama 351-0198, Japan}
\affiliation{Department of Physics, The University of Michigan, Ann Arbor, 48109-1040 Michigan, USA}

\author{Huan-Yu Ku}
\email{Huan-Yu.Ku@oeaw.ac.at}
\affiliation{Faculty  of  Physics,  University  of  Vienna,  Boltzmanngasse 5, 1090 Vienna,  Austria}
\affiliation{Institute  for  Quantum  Optics  and  Quantum  Information  (IQOQI), Austrian  Academy  of  Sciences,  Boltzmanngasse 3, 1090 Vienna,  Austria}

\author{Yueh-Nan Chen}
\email{yuehnan@mail.ncku.edu.tw}
\affiliation{Department of Physics and Center for Quantum Frontiers of Research \& Technology (QFort), National Cheng Kung University, Tainan 701, Taiwan}
\affiliation{Theoretical Quantum Physics Laboratory, Cluster for Pioneering Research, RIKEN, Wakoshi, Saitama 351-0198, Japan}

\date\today

\begin{abstract}
Quantum steering is an important correlation in quantum information theory.
A recent work [Nat. Commun. \textbf{12}, 2410 (2021)] showed that quantum steering is also useful for quantum metrology. 
Here, we extend the exploration of steering-enhanced quantum metrology from single noiseless phase shifts to superpositions of noisy phase shifts.
As concrete examples, we consider a control system that manipulates a target system to pass through a superposition of either dephased or depolarized phase shifts channels.
We show that using such superpositions of noisy phase shifts can suppress the effects of noise and improve metrology.
Furthermore, we also implemented proof-of-principle experiments for a superposition of dephased phase shifts on the IBM Quantum Experience, demonstrating a clear improvement on metrology.
\end{abstract}
\maketitle

\section{Introduction}\label{sec:intro}

Quantum theory allows one party (Alice) to remotely steer another party (Bob) by her choice of measurements.
Such a quantum phenomenon is called quantum (or Einstein–Podolsky–Rosen) steering. 
Although the concept of quantum steering was first proposed by Schr\"odinger in 1936~\cite{Schrdinger1936}, its information-theoretic description was formulated only quite recently, i.e., in 2007~\cite{PRL2007:Wiseman,Cavalcanti2016,UolaRMP2020,Xiang2022}.
Nowadays, not only many experimental realizations~\cite{Saunders2010,Bennet2012,Li2015,Wollmann2020,Deng2021,Slussarenko2022-2} of quantum steering have been demonstrated, but also various theoretical developments, such as quantum foundations~\cite{Quintino2014,PRL2014Uola,PRL2015Uola,Schmid2020typeindependent,Chen2021robustselftestingof,Ku2022test,Fadel2022entanglementoflocal}, and one-sided device-independent quantum information tasks~\cite{PRA2012:Branciard,Piani2015,PRA2016:Cavalcanti,Zhao2020,Tan2021,Bohr2022,KuPRX2022} have been proposed.

In addition to the information-theoretic formulation, Reid \textit{et al.}~\cite{RMP2009:Reid,Cavalcanti2009} investigated quantum steering from the viewpoint of the local uncertainty principle~\cite{Dressel2014PRA}.
The idea is that the complementary relations between a pair of Bob's non-commutative observables could violate the Heisenberg's limit, if the correlation shared by Alice and Bob is steerable.
In other words, the local uncertainty principle can be regarded as a criterion of steering.
Recently, Ref.~\cite{NC2021:BYadin} showed that Reid's criterion can be extended to the domain of quantum metrology~\cite{PRA2007Shaji,MA201189,IOP2014:GToth,SP2020:David,Meyer2021}, where Bob aims to estimate an unknown phase shift $\theta$ generated by a Hamiltonian $H$. 
An important result is that there exists a complementary relation between the variance of $H$ and the precision of the $\theta$ estimation quantified by the quantum Fisher information (QFI)~\cite{Chabuda2020,Fiderer2021,Zhou2021,Xu2022,Yu2022,Fallani2022,Chiribella2022arXiv}; and this result has also been demonstrated in an optical system~\cite{Gianani2022}.
This complementary relation can be regarded as not only a metrological steering inequality (MSI), but also a generalized local uncertainty relation.

The metrological steering task has so far only been investigated under a noiseless scenario, where the phase shift is generated by a perfect unitary evolution.
However, in a real experimental setup, the effects of noise are ubiquitous, such that the phase shifts could deviate from a perfect unitary and, thus, neutralize quantum advantages in metrology~\cite{Giovannetti2006,Escher2011,Giovannetti2011,Dobrza2014,yamamoto2021error}.
A typical source of noise comes from the inevitable interaction between a given system and its uncontrollable environments.
A question arises on how to mitigate the effects of these undesired interactions~\cite{Strikis2021PRXQuantum,Regula2021}.
Such a question has been addressed by applying many different methods, e.g., engineered reservoirs~\cite{myatt2000decoherence}, measurement-error mitigation~\cite{Nation2021,Strikis2021PRXQuantum}, and dynamical decoupling~\cite{Wise2021PRXQuantum}.

Recently, a novel approach, termed \emph{superposition of quantum channels}, has been used to enhance quantum capacity in communication tasks~\cite{Chiribella2019,Qt2020:Abbott,Goswami2020,PRL2021:Chiribella,Slussarenko2022,JDLin2022PRR,FJ2022PRA}.
In this framework, multiple quantum channels can be used.
Furthermore, an additional quantum control was introduced to determine which channel for the target system to pass through.
Hence, when the control system is prepared in a superposition state, the target system can go through these channels in a quantum-superposed manner.
One can take advantage of the quantum interference between these channels to alleviate the effects of noise~\cite{PRL2003:DanielK,Gisin2005,PRR2021:Rubino}.

In this paper, we consider the cases where the phase shifts are distorted by either pure dephasing noise or depolarizing noise.
In this sense, we denote the corresponding noise-distorted phase shifts as dephased and depolarized phase shifts, respectively.
Intuitively, the enhancement of the estimation precision decreases when the noise strength increases.
Furthermore, we investigate the influences of a superposition of both dephased and depolarized phase shifts by comparing different (coherent and incoherent) states of the control system.
We show that the control system in a coherent state can mitigate the noise and enhance the violation of the MSI.
Finally, we experimentally implemented a metrological steering task with a superposition of dephased phase shifts on the IBM Quantum (IBM Q) Experience~\cite{Kandala2017,GarcaPrez2020,Sun2021,Berke2022}.
Our experimental results clearly show that the enhancement of the MSI violation is due to the initial coherence of the control system.
We also provide noise simulations that take into account the inherent errors of the IBM Q device.

The rest of this paper is organized as follows.
In Sec.~\ref{sec:Metrological steering and noisy phase shift}, we review the metrological steering task proposed in Ref.~\cite{NC2021:BYadin} and extend the discussions to a scenario with a superposition of noisy phase shifts.
In Sec.~\ref{sec:Superposition of noisy phase shifts}, we formalize the concept of a superposition of noisy phase shifts, and we clearly show that its usefulness for addressing the metrological steering task.
In Sec.~\ref{sec:Experimental demonstration}, we show our experimental results obtained from the IBM Quantum Experience.
Finally, we summarize our results in Sec.~\ref{sec:conclusion}.

\section{A Metrological steering task}\label{sec:Metrological steering and noisy phase shift}
In this section, we briefly recall the steering-enhanced quantum metrology proposed in Ref.~\cite{NC2021:BYadin}.
We then extend the discussion to a scenario with a superposition of dephased (depolarized) phase shifts.

\begin{figure}[!htbp]
    \includegraphics[width=1\columnwidth]{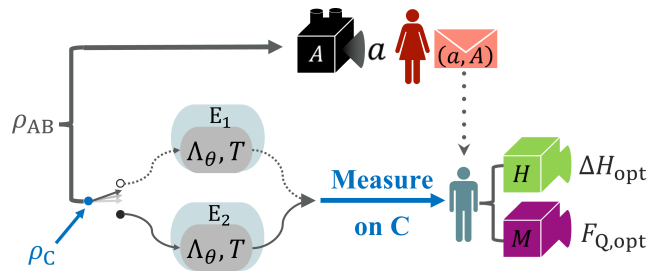}
    \caption{
    Illustration of steering-enhanced quantum metrology with a superposition of quantum channels.
    Alice (A) and Bob (B) share a bipartite state $\rho_{\A\B}$.
    Alice performs a measurement $A$ and obtains the corresponding outcome $a$.
    Then, Alice sends her information $(a,A)$ to Bob through a classical communication.
    A local phase shift $\theta$ on Bob's side is generated by a Hamiltonian $H$.
    Different from Ref.~\cite{NC2021:BYadin}, in which a phase shift is generated noiseless, we use a system C to control the evolution of system B and create a superposition of noisy phase shifts.
    When C is in the state $\ket{0}$ ($\ket{1}$), represented by the white (black) dot on the left, the system B interacts with the environment $\text{E}_1$ ($\text{E}_2$). 
    After creating the superposition of noisy phase shifts, Bob collects the conditional state and measure on C.
    According to Alice's information, Bob can decide to either measure $H$ or estimate the phase shift $\theta$ through the measurement $M$. 
    Then, he can obtain the optimal variance $\Delta H_{\opt}$ and the optimal quantum Fisher information $F_{\Q,\opt}$.}\label{fig:Superposed}
\end{figure}

We start by formulating the noiseless metrological task, where the phase shift $\theta$ is generated by a unitary channel $\exp{(-iH\theta)}$, with a ``generating" Hamiltonian $H$.
We consider a bipartite state $\rho_{\A\B}$ shared by Alice and Bob.
In each round of the experiment, Alice performs a measurement labeled by $A$.
The probability to obtain the result $a$ is denoted as $p(a|A)$; and the conditional reduced state of Bob's subsystem is $\rho_{\B,a|A}$. 
After generating a local phase shift $\theta$, Bob's conditional reduced state becomes $\rho_{\B,a|A}(\theta) = \exp(-i H\theta)\rho_{\B,a|A} \exp(iH\theta)$. 
It is convenient to summarize the result by defining an assemblage as a set of (subnormalized) quantum states, namely:
$\{\mathcal{B}_{\theta}(a,A)=p(a|A)\rho_{{\text{B}},a|A}(\theta)\}_{a,A,\theta}$.

After the measurement, Alice sends the classical information $(a,A)$ to Bob. 
Based on this information, Bob can either measure the observable $H$ or estimate the phase shift $\theta$ by measuring an observable $M$.
Note that for a given message $(a,A)$ from Alice, Bob can freely choose the observable $M$ to obtain the maximum sensitivity, quantified by the QFI $F_\Q\left(\theta|\rho_{\B,a|A}\right)$~\cite{PRL1994:Braunstein,Giovannetti2006,Giovannetti2011}. 
Here, $F_{\text{Q}}\left(\theta| \rho \right) := \Tr{L_{\theta}^2\rho(\theta)}$, where $L_\theta$ is the symmetric logarithmic derivate satisfying $\partial_{\theta}\rho(\theta)=\frac{1}{2}\{L_{\theta},\rho(\theta)\}$~\cite{IOP2014:GToth}.
The optimal QFI and the optimal variance of $H$ can be defined, respectively, as~\cite{NC2021:BYadin}:
\begin{equation}
\begin{aligned}
F_{\Q,\opt}&:=\max_{A} \sum_{a}p(a|A)F_{\Q}\left(\theta|\rho_{\B,a|A}\right), \\
\Delta H_{\opt}& :=\min_{A}\sum_{a}p(a|A)\Delta\left[\rho_{\B,a|A}(\theta),H\right]\label{Fopt},
\end{aligned}
\end{equation}
where $\Delta [\rho,H]=\Tr{H^2\rho}-\Tr{H\rho}^2$. 
Note that, in general, the QFI is evaluated for a given $\theta$~\cite{PRL2021:Tan}.

In modern terminology, the concept of local-hidden-state (LHS) model is utilized to determine whether a given assemblage is steerable or not. 
More specifically, an assemblage that admits a local-hidden-state model can be described as~\cite{PRL2007:Wiseman} 
\begin{equation}\label{eq:LHS}
\mathcal{B}^{\text{LHS}}_{\theta}(a,A)=\sum_{\lambda}p(\lambda)p(a|A,\lambda)\rho_{\B,\lambda}(\theta)\quad \forall a,A,
\end{equation}
where $\{\rho_{\B,\lambda}(\theta)\}_{\lambda,\theta}$ are quantum states and $\{p(a|A,\lambda)\}_{\lambda}$ constitute a stochastic map, which maps the hidden variable $\lambda$ into $a|A$.
If a given assemblage can be simulated by a local-hidden-state model, it is unsteerable. Otherwise, it is steerable. 
As reported in Ref.~\cite{NC2021:BYadin}, when an assemblage is unsteerable, the MSI can be derived as $F_{\Q,\opt}\leq 4\Delta H_{\opt}$.
Here, we define the violation $V$ of the MSI, i.e.,
\begin{equation}
V:=\max \left( F_{\Q,\opt} - 4\Delta H_{\opt},~0 \right).
\label{eq:Steering ineq}
\end{equation}
Therefore, $V>0$ implies that the assemblage is steerable.

\begin{table}
    \begin{tabular}{ccccc}
      \hline\hline
      
      $A$ & \multicolumn{2}{c}{$\sigma_x$} & \multicolumn{2}{c}{$\sigma_z$} \\[3pt]
      \hline
      $a$ & $0$ & $1$ & $0$ & $1$\\[2pt]

      $p(a|A)$ & $0.5$ & $0.5$ & $0.5$ & $0.5$\\[2pt]
      
      $~~\rho_{a|A}~~$ & $~~\proj{+}~~$ & $~~\proj{-}~~$ & $~~\proj{0}~~$ & $~~\proj{1}~~$ \\[2pt]
      
      \hline\hline
    \end{tabular}\caption{Summary of results for Alice's measurements $A$ with outcomes $a$, which include the probability $p(a|A)$ and the corresponding post-measured state $\rho_{a|A}$.\label{tab:Stateprepare}}
\end{table}

\section{A Superposition of noisy phase shifts}\label{sec:Superposition of noisy phase shifts}
Throughout this paper, we consider that a noisy phase shift channel $\Lambda_{\theta}$ can be described by a noiseless one followed by a noisy channel $\Lambda$~\cite{Escher2011}, i.e.,
\begin{equation}
    \Lambda_{\theta}(\rho) = \Lambda\left( e^{-iH\theta}~ \rho~ e^{iH\theta}\right),\label{eq:Noisy evolution}
\end{equation}
where the noisy channel $\Lambda$ commutes with the unitary $U=\exp{(-iH\theta)}$, i.e., $\Lambda(U\rho U^{\dag}) = U\Lambda(\rho)U^{\dag}$, to guarantee that $H$ is still the generating Hamiltonian of $\theta$ in the output noisy states.
We now consider a scenario for superposing two identical noisy phase shifts, as shown in \fig{\ref{fig:Superposed}}.
According to Ref.~\cite{Qt2020:Abbott}, a superposition of multiple channels is well-defined if the implementation of each member channel is specified.
More specifically, according to the Stinespring dilation theorem~\cite{Stinespring1955,Kraus1983,Wilde2017}, there exist non-unique system-environment models to describe the channel $\Lambda_\theta$, namely
\begin{equation}
\begin{aligned}
\exists U_{\text{BE}}(\theta),~\mathcal{E}_{\text{E}}~\text{s.t.}~ \Lambda_{\theta}(\rho)&=\text{Tr}_{\text{E}}[U_{\text{BE}}(\theta)(\rho \ts \mathcal{E}_{\text{E}})U_{\text{BE}}^{\dag}(\theta)],
\label{eq:Unitary to Kraus operator}
\end{aligned}
\end{equation}
where $U_{\text{BE}}(\theta)$ denotes the system-environment global unitary, and $\mathcal{E}_{\text{E}}$ is the initial state of the environment.
Here, we introduce a quantum control system C to determine which environment (i.e., $\text{E}_1$ or $\text{E}_2$,) affects the system B.
We consider that the total system is initially prepared in 
\begin{equation}
    \rho_{\text{tot}} = |j\rangle \langle j|_{\text{C}} \otimes \rho \otimes \mathcal{E}_{\text{E}_1} \otimes \mathcal{E}_{\text{E}_2},
\end{equation}
for $j$ being either $0$ or $1$.
In this case, the total evolution can be described by 
\begin{equation}
U_{\text{tot}}=\ket{0}\bra{0}_{\text{C}} \ts~U_{\text{BE}_1}(\theta) +\ket{1}\bra{1}_{\text{C}} \ts~U_{\text{BE}_2}(\theta).\label{eq:Total unitraty}
\end{equation}
The reduced state of C and B reads
\begin{align}
\rho_{\C\B}(\theta)&=\text{Tr}_{\text{E}_{1},\text{E}_{2}}\left[U_{\text{tot}}\left(|j\rangle \langle j|_{\text{C}}\ts \rho \ts \mathcal{E}_{\text{E}_1} \ts \mathcal{E}_{\text{E}_2}\right)U_{\text{tot}}^{\dagger}\right] \nonumber\\
&= |j\rangle \langle j|_{\text{C}} \ts \text{Tr}_{\text{E}_j}[U_{\B\text{E}_j}(\rho \ts \mathcal{E}_{\text{E}_j})U_{\text{BE}_j}^{\dagger}] \nonumber\\
&= |j\rangle \langle j|_{\text{C}} \otimes \Lambda_\theta (\rho).\label{eq:mixstate control}
\end{align}
In other words, when C is prepared in the state $|j\rangle$, B interacts with the corresponding environment $\text{E}_j$.
Thus, if C is prepared in an incoherent mixed state, i.e., $ (\ket{0}\bra{0}_{\text{C}}+\ket{1}\bra{1}_{\text{C}})/2$, the system B has equal probabilities to interact with either $\text{E}_0$ or $\text{E}_1$.
For simplicity, we consider that $U_{\text{BE}_1}(\theta)$ and $U_{\text{BE}_2}(\theta)$ are isomorphic to each other (so $\mathcal{E}_{\text{E}_1}$ = $\mathcal{E}_{\text{E}_2}$); that is, two phase shifts are implemented in the same way.

On the other hand, when the control C is prepared in $\ket{+}_{\text{C}} = (\ket{0}_{\text{C}}+\ket{1}_{\text{C}})/\sqrt{2}$, we obtain
\begin{equation}
\rho_{\C\B}(\theta)
= \frac{\id_{\text{C}}}{2}\ts \Lambda_{\theta}(\rho)
+\frac{(\ket{0}\bra{1}_{\text{C}}+\ket{1}\bra{0}_{\text{C}})}{2}\ts  T\rho~T^{\dagger},\label{eq:Transformation} 
\end{equation}
where $T=\text{Tr}_{\text{E}}\left[U_{\text{BE}}~(\id \ts\mathcal{E})\right]$ characterizes the quantum interference effect between these two channels~\cite{Qt2020:Abbott}.
The interference effect occurs simultaneously with the non-zero off-diagonal terms in C.
In this case, the target passes through a ``superposition of noisy phase shift channels".
Note that we have omitted the subscripts for the environments because they are isomorphic to each other.

Now, we perform a set of projective measurements, $\left\{\ket{+}\bra{+}_{\C}, \ket{-}\bra{-}_{\C}\right\}$, with $|\pm\rangle = (\ket{0}\pm \ket{1})/\sqrt{2}$, on the quantum control C.
The post-measured states of B then read
\begin{equation}
\begin{aligned}
\rho_{\B,\pm}(\theta)
&=\frac{\text{Tr}_\C \left[(\ket{\pm}\bra{\pm}_{\text{C}} \ts \id_{\B})~\rho_{\C\B}(\theta)\right]}{\Tr{(\ket{\pm}\bra{\pm}_{\text{C}} \ts \id_{\B})~\rho_{\C\B}(\theta)}}\\
&=\frac{\Lambda_{\theta}(\rho)\pm T\rho~T^{\dagger}}{2P_\pm},\label{eq:Post-Measure}
\end{aligned}
\end{equation}
where $P_\pm=\Tr{(\ket{\pm}\bra{\pm}_{\text{C}} \ts \id_{\B})~\rho_{\C\B}(\theta)}$ are the probabilities of the outcomes $\pm$ for the projective measurements. 
Equation~\eqref{eq:Post-Measure} shows that the post-measured state does not only depend on the noisy phase shift $\Lambda_\theta$, but also on the quantum interference effects described by $T$.

We are now ready to demonstrate the main result of this paper that: the superposition of phase shifts can enhance the violation of the MSI.
To highlight this point, we compare the two cases:
\begin{enumerate}[label=(\arabic*)]
    \item control C is prepared in an incoherent mixed state $\id_{\text{C}}/2$ (\emph{without} a superposition of phase shifts),
    \item control C is in a superposition $\ket{+}\bra{+}_{\text{C}}$ state (\emph{with} a superposition of phase shifts).
\end{enumerate}
We show that case (1), in general, cannot improve the violation of MSI; nevertheless, for case (2), it is possible to observe an enhancement of the MSI violation under some circumstances.

Let us now investigate the post-measured states to gain more insight.
For case (1), we can observe that the reduced state of C and B is separable, i.e., $\rho_{\text{CB}}=\id_{\text{C}}/2 \otimes \Lambda_\theta (\rho)$, and thus, the measurement $\ket{\pm}\bra{\pm}_\C \ts \id_{\B}$ on this separated state cannot affect the system B.

After tracing out the control system C, we observe that the post-measured state is $\Lambda_\theta(\rho)$, which is identical to using a single-noise phase shift.
In this case, the violation cannot be enhanced [see the task discussed in Eq.~\eqref{eq:avg FI and V}], because both optimal QFIs (variances) calculated from the post-measured state, i.e., $F_{\text{Q,opt},\pm}$ $(\Delta H_{\text{opt},\pm})$ are the same as the original QFI (variance) in Eq.~\eqref{Fopt}.
Thus, case (1) cannot improve our task for any kind of noisy phase shifts.
Note that although we only consider the maximally mixed state, this result generally holds for all convex mixtures of the states $\ket{0}\bra{0}_{\text{C}}$ and $\ket{1}\bra{1}_{\text{C}}$.

For case (2), we consider two concrete examples: the dephased and depolarized phase shifts are respectively characterized by the following system-environment unitary evolution:
\begin{equation}
\begin{aligned}
U^{\text{deph}}_w\ket{\psi}\ts \ket{0}_{\text{E}} &= \sqrt{1-\frac{w}{2}}~\ket{\psi_{\theta}}\ts \ket{0}_{\text{E}}\\
&~+\sqrt{\frac{w}{2}}~\sigma_z \ket{\psi_{\theta}}\ts \ket{1}_{\text{E}},\label{Sim:Depahsing Unitary}
\end{aligned}
\end{equation}
\begin{equation}
\begin{aligned}
U^{\text{depo}}_v \ket{\psi}\ts \ket{0}_{\text{E}} &= \sqrt{1-\frac{3v}{4}}\ket{\psi_{\theta}}\ts \ket{0}_{\text{E}}\\
&~+\sqrt{\frac{v}{4}} \sigma_x \ket{\psi_{\theta}}\ts \ket{1}_{\text{E}}\\
&~+\sqrt{\frac{v}{4}} \sigma_y \ket{\psi_{\theta}}\ts \ket{2}_{\text{E}}\\
&~+\sqrt{\frac{v}{4}} \sigma_z \ket{\psi_{\theta}}\ts \ket{3}_{\text{E}},\label{Sim:Depolarizing Unitary}
\end{aligned}
\end{equation}
where $\ket{\psi_{\theta}} = \exp{(-iZ\theta)} \ket{\psi}$, and $w$ ($v$) is the visibility for the dephased (depolarized) phase shift.

The post-measured states conditioned on the results ‘‘$\pm$" and, according to Eq.~\eqref{eq:Post-Measure}, can be written as:
\begin{equation}
\begin{aligned}
\rho^{\text{deph}}_{\B,a|A,\pm}(\theta)&=\frac{\Lambda^{\text{deph}}_{\theta,w}(\rho_{\B,a|A})\pm (1-\frac{w}{2})\rho_{\B,a|A}(\theta)}{1\pm (1 -\frac{w}{2})},\\
\rho^{\text{depo}}_{\B,a|A,\pm}(\theta)&=
\frac{\Lambda^{\text{depo}}_{\theta,v}(\rho_{\B,a|A})\pm\left(1-\frac{3v}{4}\right) \rho_{\B,a|A}(\theta) }{1\pm (1-\frac{3v}{4})},
\end{aligned}
\end{equation}
where $\rho_{\B,a|A}(\theta)=\exp{(-iZ\theta)}\rho_{\B,a|A}\exp{(iZ\theta)}$, and $\Lambda^{\text{deph}}_{\theta,w}$ ($\Lambda^{\text{depo}}_{\theta,v}$) is denoted as a single-use of the dephased (depolarized) phase shift.
Additionally, the probabilities of the dephased and depolarized phase shifts are:
\begin{equation}
\begin{aligned}
    P^{\text{deph}}_{\pm}& =  \frac{1}{2} \pm (\frac{1}{2}-\frac{w}{4}) \\
    P^{\text{depo}}_{\pm}& =  \frac{1}{2} \pm (\frac{1}{2}-\frac{3v}{8}).
\end{aligned}
\end{equation}

One can discover that the post-measured state with $\rho_{\text{C}}=\ket{+}\bra{+}_{\text{C}}$ can be effectively characterized by a mixture of a noisy phase shift, i.e., $\Lambda^{\text{deph}}_{\theta,w}(\rho_{\B,a|A})$ or $\Lambda^{\text{depo}}_{\theta,v}(\rho_{\B,a|A})$, and a noiseless shift, i.e., $\rho_{\B,a|A}(\theta)$.
Thus, the effects of noise can be probabilistically decreased.

For dephased phase shifts, the post-measured states on ``+" can be seen as a state suffering from another dephased noise with visibility $w'$; that is
\begin{equation}
\begin{aligned}
\rho^{\text{deph}}_{\B,a|A,+}(\theta)
&=\frac{\Lambda^{\text{deph}}_{\theta,w}(\rho_{\B,a|A})+(1-\frac{w}{2})\rho_{\B,a|A}(\theta)}{2-\frac{w}{2}}\\
&=\frac{2(1-\frac{w}{2})\rho_{\B,a|A}(\theta) + \frac{w}{2}\sigma_z \rho_{\B,a|A}(\theta) \sigma_z }{2-\frac{w}{2}}\\
&=\left(1- \frac{w'}{2} \right)\rho_{\B,a|A}(\theta)+\frac{w'}{2} \sigma_z \rho_{\B,a|A}(\theta) \sigma_z,
\end{aligned}
\end{equation}
where $w'=2w/(4-w)$.
One can observe that $w'<w$ when $w \in [0,1]$, which indicates that the effect of noise can be mitigated. For the ``$-$" case, we observe that the post-measured state undergoes the unitary transform $\sigma_z$ and is independent of $w$, i.e., $\rho^{\text{deph}}_{\B,a|A,-}(\theta)=\sigma_z \rho_{\B,a|A}(\theta) \sigma_z$.
Both ``$\pm$" cases of the post-measured states include the information of the unknown phase shift $\theta$. 

To further discuss the coherent-control-enhanced violation of the MSI, we consider the average optimal QFI and variance by taking into account their probabilities~\cite{SP2020:David}, namely:
\begin{equation}
F_{\text{Q,opt}}^{\text{avg}}:=\sum_{\pm} P_{\pm} F_{\text{Q,opt},\pm}; \quad  \Delta H_{\text{opt}}^{\text{avg}}:=\sum_{\pm} P_{\pm}\Delta H_{\text{opt},\pm}.\label{eq:avg FI and V}
\end{equation}

\begin{figure}[!htbp] 
    \subfloat{\includegraphics[width=.99\linewidth]{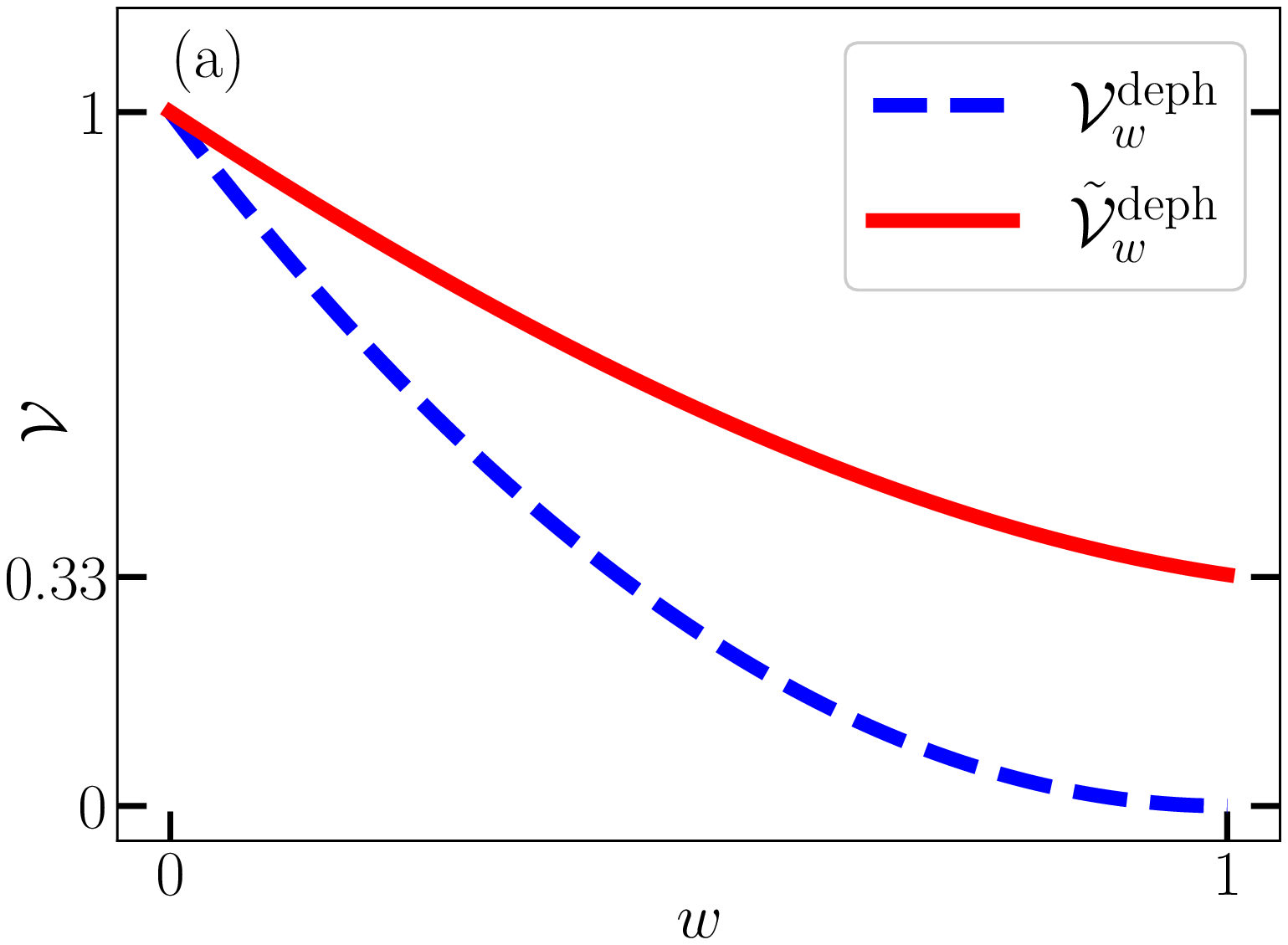}\label{fig:Thmsol_pha}}
    \vspace{0.01\linewidth}
    \subfloat{\includegraphics[width=.99\linewidth]{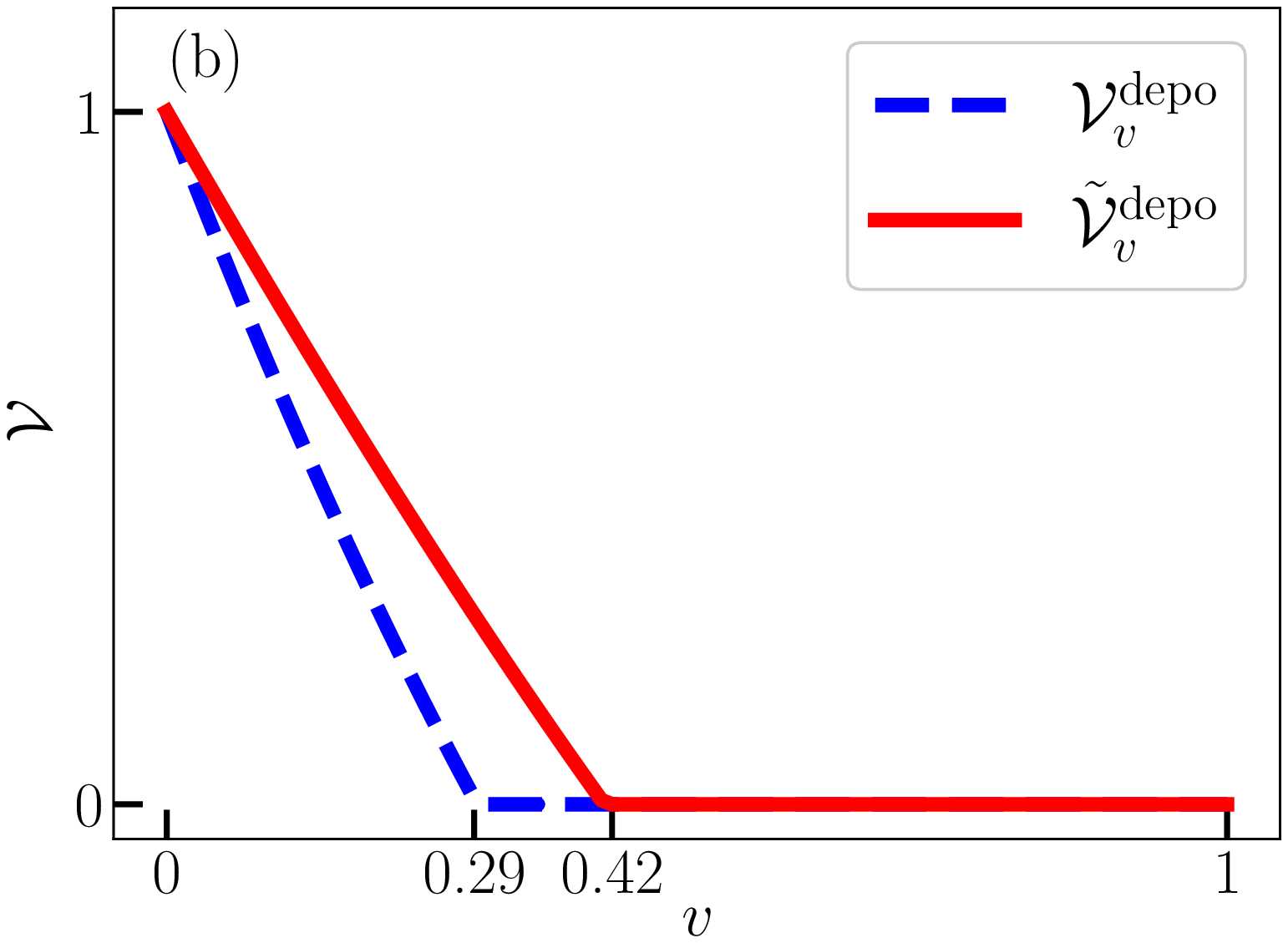}\label{fig:Thmsol_pol}}
    \caption{
    Average violations $\mathcal{V}$ of the metrological steering inequality. 
    We set $\theta=0$ and plot the average violations of two examples: (a) dephased and (b) depolarized phase shifts with the visibilities $w$ and $v$, respectively.
    Note that $\rho_{\text{C}}=\id_{\text{C}}/2$ and $\rho_{\text{C}}=\ket{+}\bra{+}_{\text{C}}$ represent different choices of the initial states of the control systems.
    One can observe in example (a) that the control $\rho_{\text{C}}=\ket{+}\bra{+}_{\text{C}}$ can enhance the violation for a given visibility; although the system is fully dephased, we can still witness $\mathcal{V} \approx 0.33$ with $\rho_{\text{C}}=\ket{+}\bra{+}_{\text{C}}$.
    For example (b), the depolarized noise (with $\rho_{\text{C}}=\id_{\text{C}}/2$) causes a sudden-vanishing of the violation when $v\approx0.29$; however, if $\rho_{\text{C}}=\ket{+}\bra{+}_{\text{C}}$, it can enhance the violation and extend the sudden-vanishing effect to $v\approx 0.42$.
    \label{fig:Thmsol}}
\end{figure}

In Fig.~\ref{fig:Thmsol}, we present the average violations $\mathcal{V}$ of MSI, i.e.,
\begin{equation}
   \mathcal{V}=\max(F_{\text{Q,opt}}^{\text{avg}}-4\Delta H_{\text{opt}}^{\text{avg}},~0),
\end{equation}
of the two phase shifts, in which the Bell state $\ket{\Phi^{+}}_{\text{AB}}=(\ket{00}+\ket{11})/\sqrt{2}$ and the Pauli observable $A=\{ \sigma_x,\sigma_z\}$ are considered.
We denote the average violations of the two examples as:
\begin{enumerate}[label=(\alph*)]
    \item dephased phase shifts with visibility $w$ labeled by $\mathcal{V}^{\text{deph}}_w$ ($\tilde{\mathcal{V}}^{\text{deph}}_w$),
    \item depolarized phase shifts with visibility $v$ denoted by $\mathcal{V}^{\text{depo}}_v$ ($\tilde{\mathcal{V}}^{\text{depo}}_v$)
\end{enumerate}
with C initially prepared in $\id_{\text{C}}/2$ $\left(\ket{+}\bra{+}_{\text{C}}\right)$, respectively.

For the example of the dephased phase shifts, as shown in \fig{\ref{fig:Thmsol}}\red{(a)}, one can observe that the system with a superposition of dephased phase shifts has a clear enhancement of the violations for a given visibility $w$ [see Fig.~\ref{fig:Thmsol}\red{(a)}].
Remarkably, though the system is completely dephased ($w=1$), we can still find $\mathcal{V} \approx 0.33$.

For the example of depolarized phase shifts, the superposition of depolarized phase shifts can enhance the violation and extend the sudden-vanishing of the violation from $v\approx0.29$ to $v\approx 0.42$ [see also Fig.~\ref{fig:Thmsol}\red{(b)}].

Note that one can consider a more general coherent state for the control qubit, i.e., 
$\ket{\psi}_{\text{C}} = \sqrt{\alpha}\ket{0}_{\text{C}} + \sqrt{1-\alpha}\ket{1}_{\text{C}}$, where $\alpha\in [0,1]$ determines the degree of coherence of the state. More specifically, the degree of coherence vanishes when either $\alpha = 0$ or $\alpha=1$, and it monotonically increases (decreases) in the region $[0,0.5]$ ([0.5,1]). One can further find that the degree of the methodological enhancement agrees with the amount of the system C's initial coherence.

\section{Experimental demonstration}\label{sec:Experimental demonstration}

In this section, we propose a circuit model of superposition of dephased phase shift that only consists of 12 CNOT gates and 17 single-qubit gates, and demonstrate the enhancement on a IBM Q processor.
Additionally, we simulate the device-intrinsic noise to identify the effects of noise in our experimental data.

To further decrease the circuit depth, we consider a scenario known as temporal steering~\cite{PRA2014:YNChen,Bartkiewicz2016,PRL2016:SLChen,Bartkiewicz2016Scireport}.
Therein, the initial maximally entangled state shared by Alice and Bob can be replaced by a prepare-and-measure scenario~\cite{Li2015PRA,Armin2021PRXQuantum}. 
More specifically, under the temporal steering scenario, Alice now measures $\sigma_x$ and $\sigma_z$ on the maximally mixed state $\id/2$, instead of performing local measurements on the bipartite state $\ket{\psi}_{\A\B}$.
Note that since the IBM Q does not allow its users to manipulate the post-measured state, we directly prepare the eigenstates of $\sigma_x$ and $\sigma_z$ with probability $p(a|A)$. 
In this way, one can obtain the same assemblage as in Table~\ref{tab:Stateprepare}, before we start the noisy metrological test.

After the initial assemblage is constructed, there is no operational difference between spatial and temporal steering in the metrological test, because the property of the maximally entangled state $X\otimes Y\ket{\Phi^{+}}_{\text{AB}}=\openone\otimes YX^T\ket{\Phi^{+}}_{\text{AB}}$.
Thus, we only focus on Bob's subsystem (see also the similar discussion in Ref.~\cite{KuPRX2022}).   
Under the assumptions of macrorealism~\cite{Leggett85,Emary14} (i.e., the properties of the system are well defined and measurements do not disturb the system), a temporally classical assemblage can be expressed by a hidden-state model described in the same form of local-hidden-state model c.f., Eq.~\eqref{eq:LHS}. 
In other words, if the hidden-state model is satisfied, the noisy channel breaks the temporal steerability such that the collections of states are well defined.

\subsection{Circuit implementation on the IBM Q}
\begin{figure*}[!htbp]
    \includegraphics[width=1.99\columnwidth]{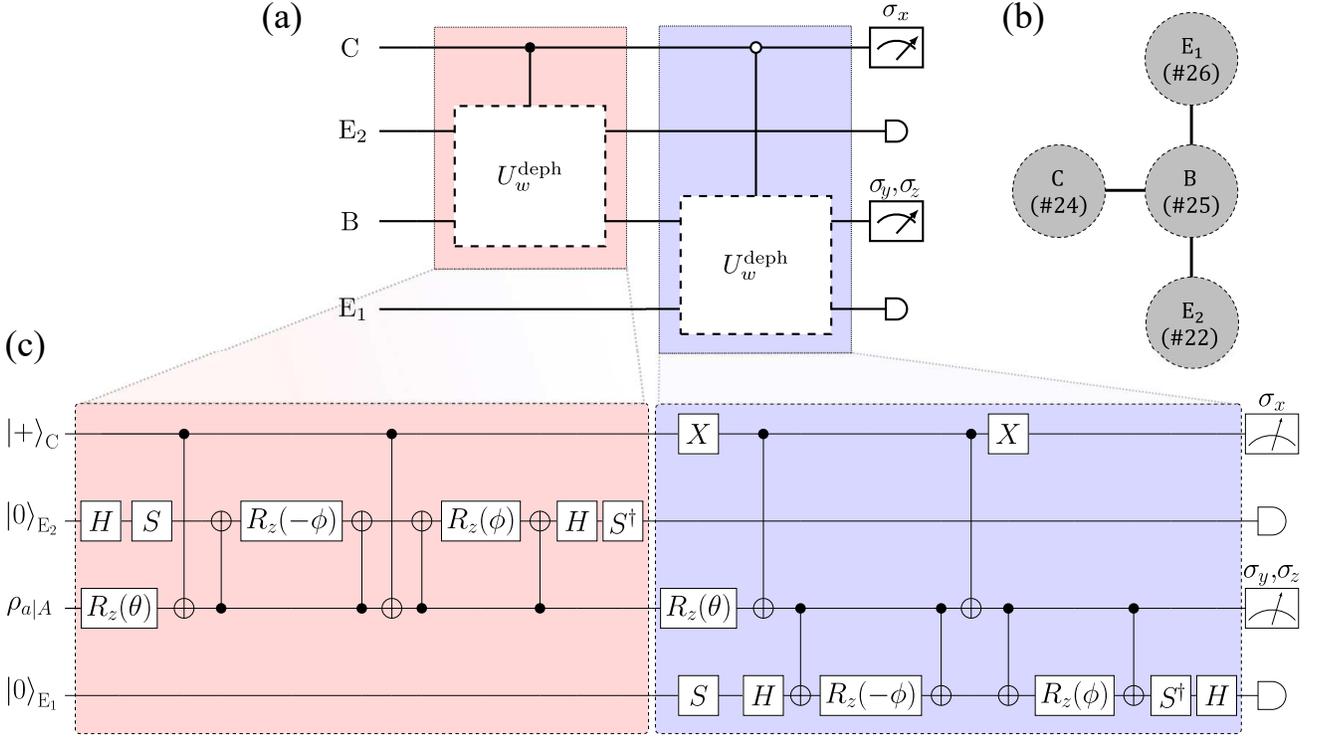}
    \caption{Circuit model for steering-enhanced quantum metrology with a superposition of dephased phase shifts. 
    Schematics of our circuit model (a) without and (c) with details. (b) In the topology of the four qubits that we chose in the IBM-Cairo device, the numbers label the qubits \#24, \#25, \#26, \#22, representing the system C, B, $\text{E}_{1}$, $\text{E}_{2}$, respectively. 
    Here, the visibility $w$ is tuned by the angle $\phi$, such that $w=2\sin^2 (\phi/2)$. We use the standard symbols for quantum gates (see Appendix A).\label{fig:IBM_Expsetting}}
\end{figure*}

As shown in Fig.~\ref{fig:IBM_Expsetting}, we provide a circuit model to experimentally implement the metrological steering task with the superposition of dephased phase shifts described in the previous section [Eqs.~\eqref{eq:Total unitraty} and~\eqref{Sim:Depahsing Unitary}].
This circuit involves four qubits, which serve as the control C, the system B, and the two environments, $\text{E}_1$ and $\text{E}_2$, respectively.
Because CNOT gates on the IBM Q are restricted by the connectivity of the devices, we find that the implementation of the circuit on the devices with the coupling map shown in Fig.~\ref{fig:IBM_Expsetting}\red{(b)} can minimize the number of CNOT gates.

This circuit can be divided into three parts: (i) state preparation, (ii) the superposition of dephased phase shifts, and (iii) measurement on the qubits C and B.
In part (i), the qubits C, B, and $\text{E}_{1,(2)}$ are prepared in the states $\proj{+}$, $\rho_{a|A}$, and $\proj{0}_{1,(2)}$, respectively.
In the IBM Q device, all qubits are initially in the state $\ket{0}$. The state preparation can be achieved by applying single-qubit gates on each qubit. 
For instance, we can obtain a $\ket{+}_{\text{C}}$ by applying a Hadamard gate on the control system C.

In part (ii), the circuit model of the superposition of dephased phase shifts is shown in Fig.~\ref{fig:IBM_Expsetting}\red{(a)}.
The qubit topology of the four qubits that we chose in IBM-Cairo is shown in Fig.~\ref{fig:IBM_Expsetting}\red{(b)}.
Through the control qubit C, the system B can interact with alternative environments.
We divide the total unitary in Fig.~\ref{fig:IBM_Expsetting}\red{(a)} into a gate sequence, which is shown in Fig.~\ref{fig:IBM_Expsetting}\red{(c)}.
In this sequence, we use control-rotation with angle $\phi$ on the system B and its corresponding environment. 
After we trace out its environment, this control-rotation gate is effectively equal to the pure dephasing noise on the system B.
Here, the visibility of the pure dephasing noise $w$ is tuned by the angle $\phi$, such that $\phi=2\sin^{-1}(\sqrt{w/2})$, with $\phi \in [0,\pi/2]$.
In part (iii), we measure $\sigma_x$ on qubit C and measure $\sigma_z$ or $\sigma_y$ on qubit B. 
Note that IBM Q only allows us to conduct $\sigma_z$ measurements. 
Therefore, we can apply a Hadamard gate ($H$) before $\sigma_z$ measurement to obtain $\sigma_x$, and a phase gate ($S$) plus Hadamard gate to obtain the measurement $\sigma_y$.

Let us now elaborate how to obtain the Fisher information (FI) and the variance from the measurement results.
The measurement data can be summarized by a set of probabilities $\{p_{\theta,\phi}(b,c|M,\rho_{a|A,c})\}$, where $M\in \{\sigma_z,\sigma_y\}$ denotes Bob's measurement with the outcome $b\in \{0,1\}$, and $c\in \{+,-\}$ is the outcome of measuring $\sigma_x$ on C.
Note that $\{M_b\}_b$ is the set of positive operators that satisfy $\sum_b M_b = \id$. The probability $p(b|M)$ is given by Born's rule, i.e., $p(b|M,\rho)=\text{Tr}[M_b \rho ]$.
The marginal probabilities then read
\begin{equation}
    p_{\theta,\phi}(b|M,\rho_{a|A})=\sum_c p_{\theta,\phi}(b,c|M,\rho_{a|A,c}).
\end{equation}
In addition, the optimal FI reads
\begin{equation}\label{eq:conditionalFI}
F_{\opt,\pm}:= \max_A \sum_{a}p(a|A,\pm)F\left(\theta |M,\rho_{a|A,\pm}\right),
\end{equation}
where $F\left(\theta |M,\rho \right)$ denotes the FI obtained from a state $\rho$ with measurement $M$, which is defined as
\begin{equation}\label{eq:FI}
F\left(\theta |M,\rho \right):= \sum_{b}\frac{\left[\partial_{\theta} p_{\theta}(b|M,\rho)\right]^{2}}{p_{\theta}(b|M,\rho)}.
\end{equation}
Note that the FI for a given measurement $M$ is a lower bound of QFI i.e., $ F\left(\theta |M,\rho \right) \leq F_\Q(\theta|\rho )$~\cite{IOP2014:GToth}, and thus, $F_{\opt}\leq F_{\Q,\opt}$.

As a similar approach in Eq.~\eqref{eq:avg FI and V}, we take both outcomes $c=+$ and $c=-$ with the probabilities $P_{\pm}$ into account to obtain the average optimal FI, i.e.,
\begin{equation}
    F_{\text{opt}}^{\text{avg}}:=\sum_{\pm} P_{\pm} F_{\text{opt},\pm}.\label{eq:avgexam}
\end{equation}

\begin{figure*}[!htbp]
    \subfloat{\includegraphics[width=.99\columnwidth]{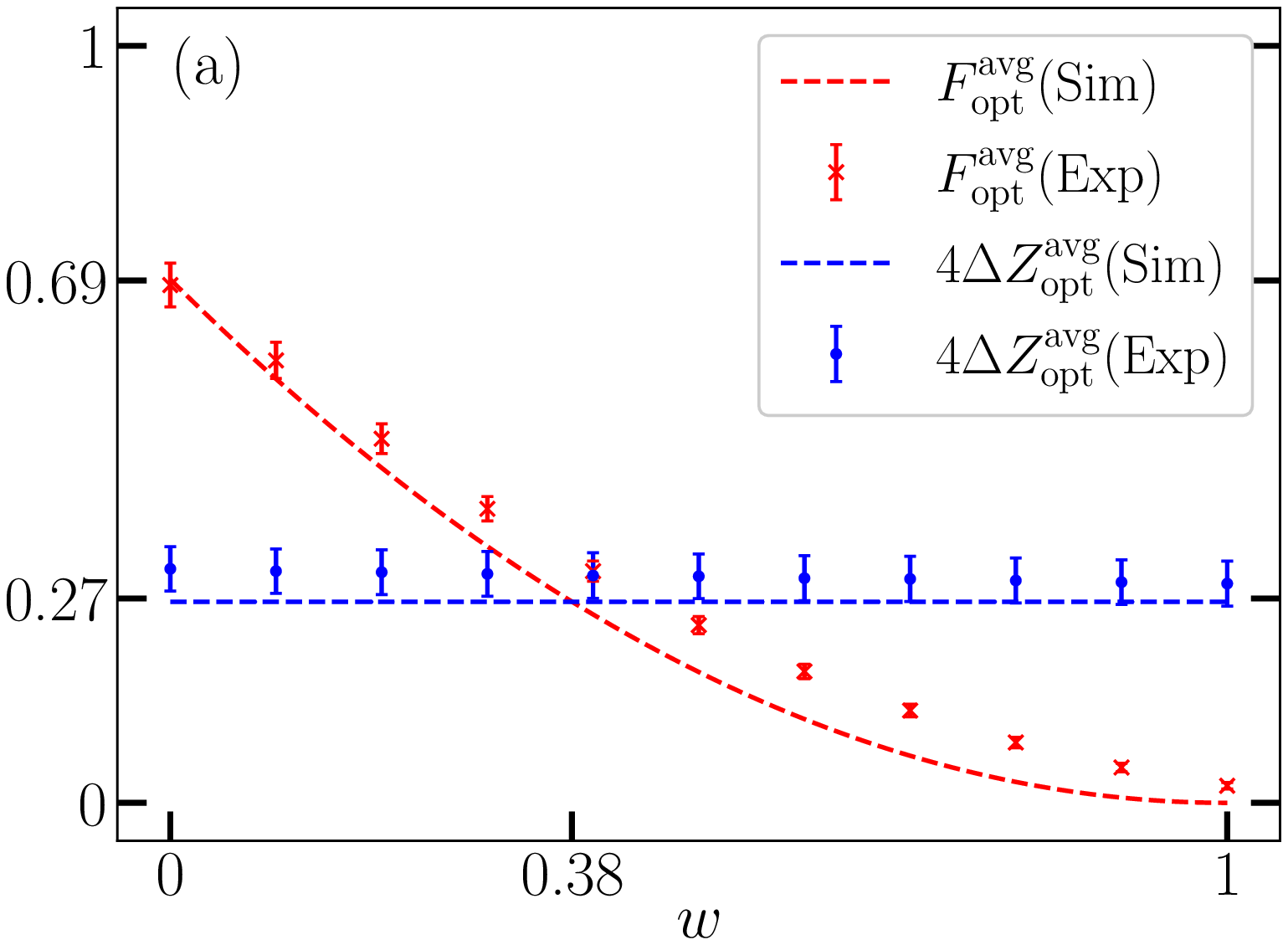}}
    \hspace{0.5cm}
    \subfloat{\includegraphics[width=.9915\columnwidth]{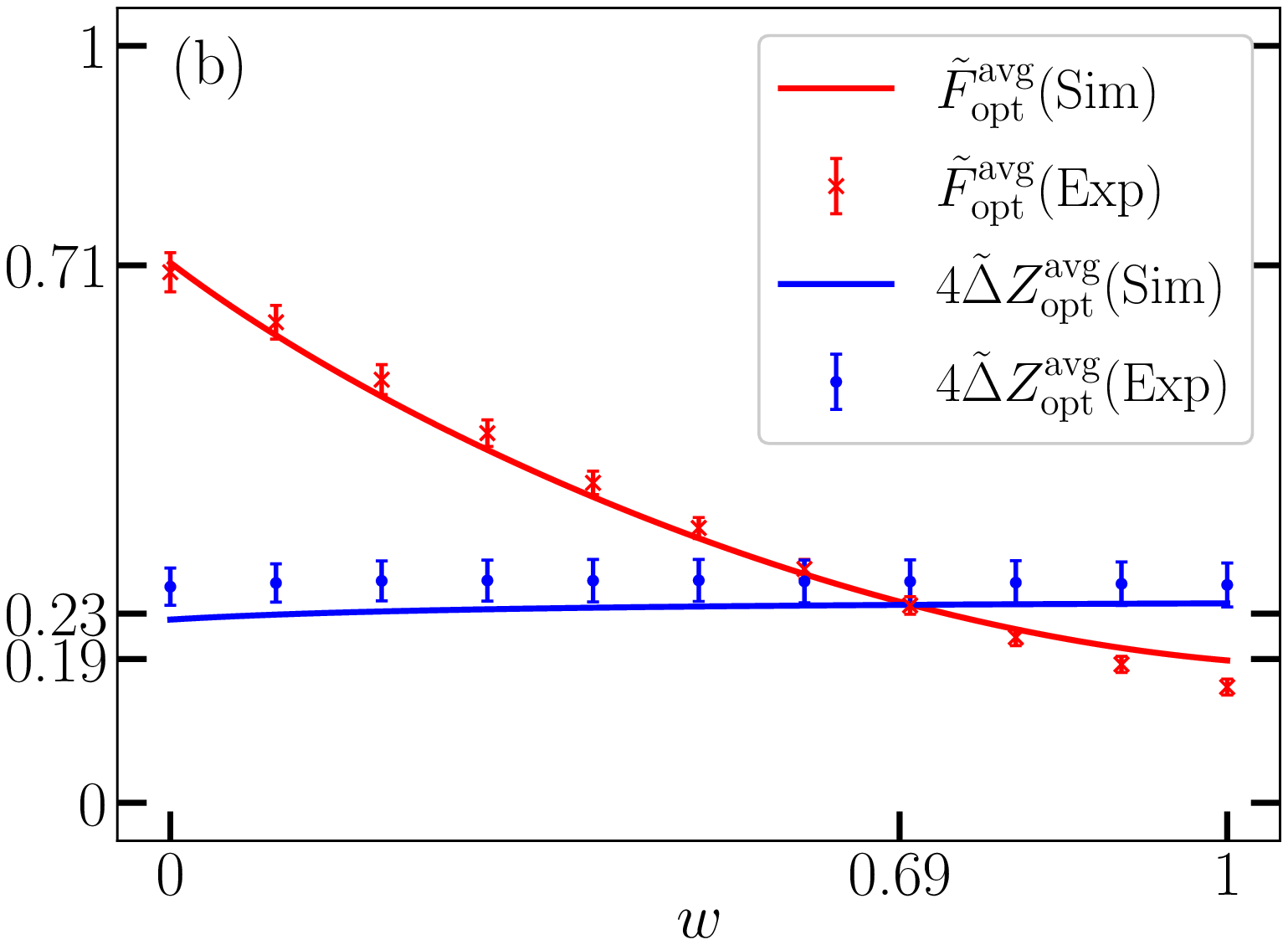}}
    \caption{
    Experimental results and noise simulations of the metrological tasks with the dephased phase shifts: the control C is prepared in (a) $\rho_{\text{C}}=\id_{\text{C}}/2$ (\emph{without} a superposition of phase shifts) and (b) $\rho_{\text{C}}=\ket{+}\bra{+}_{\text{C}}$ (\emph{with} a superposition of phase shifts). 
    Specifically, the red-cross (blue-circle) data points are the experimental results of the average optimal Fisher information (the optimal variance) with respect to the visibility $w$.
    Note that $\theta=0$. 
    Here, the error bars are obtained from $40$ individual rounds of experiments; each experimental data point consists of 10,000 individual runs performed on about ten different dates. Therefore, the error bars represent the variance of the IBM-Cairo device to conduct these experiments in long timescales. 
    The solid curves represent the noise simulations for the dephased phase shifts with $\rho_{\text{C}}=\ket{+}\bra{+}_{\text{C}}$; the dashed curves represent the noise simulations with $\rho_{\text{C}}=\id_{\text{C}}/2$. 
    Although we consider only several common noisy resources, the tendencies and magnitudes of noise simulations approach the actual experiment in both cases (a) and (b). 
    We clearly observe that the control in a superposition state, i.e., $\ket{+}\bra{+}_{\text{C}}$, can enhance the optimal Fisher information and decrease the optimal variance; thus, it extends the violations of the metrological steering inequality from $w \approx 0.38$ to $w \approx 0.69$.\label{fig:IBM_Cairo_result}}
\end{figure*}

We implement our proposal on the IBM-Cairo device because it has longer relaxation and coherence times, i.e., $T_1$, $T_2$, and lower gate errors than other available IBM Q devices~(see Table~\ref{Tab:IBM Q} and the information from IBM Q website~\cite{IBMWeb}).
In addition, we choose the qubits, labeled by \#25, \#24, \#26, and \#22 in the device, to represent B, C, $\text{E}_1$, and $\text{E}_2$, respectively, because of their connectivities [see \fig{\ref{fig:IBM_Expsetting}\red{(b)}}]. 
As shown in Fig.~\ref{fig:IBM_Cairo_result}, we provide the results by conducting experiments on the IBM-Cairo device with $10,000$ shots for each data point.

To calculate the partial derivative of the probability in Eq.~\eqref{eq:FI}, we use a fitting function $g(\theta)= 0.5+\alpha \sin(2\theta+\beta)$, to interpolate the $p_{\theta,\phi}$, where $\alpha$ and $\beta$ are fitting parameters.
Also, we take $\theta=0$ to obtain the maximum value of the optimal FI. 
We observe that the system with a control state $\rho_{\text{C}}=\ket{+}\bra{+}_{\text{C}}$ can increase $F_{\opt}^{\text{avg}}$ and decrease $\Delta Z_{\text{opt}}^{\text{avg}}$.
The threshold of the average MSI violations can be increased, i.e., from $w \approx 0.38$ to $w \approx 0.69$. 

\begin{figure}[!htbp] 
    \centering
    \includegraphics[width=0.99\linewidth]{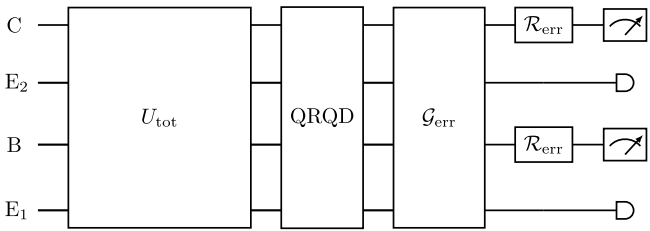}
    \caption{
    Model of noise simulations. We model the qubit relaxation and qubit dephasing effects that occur after performing a total unitary evolution. 
    We apply depolarizing and bit-flip channels to describe gate errors and read-out errors (on the system B and C), respectively.
    Here, QRQD represents the qubit relaxation and qubit dephasing channel which action is given by the master equation in Eq.~\eqref{eq:Mastereq}; the CNOT error $\mathcal{G}_{\text{err}}$ is modelled by the depolarizing noise in Eq.~\eqref{eq:CNOTErr}; the readout error $\mathcal{R}_{\text{err}}$ is described by the bit-flip channel in Eq.~\eqref{eq:ReadoutErr}.\label{fig:NoiseModel}}
\end{figure}
\subsection{Noise simulations}
Here, we also provide noise simulations by using NumPy and QuTip~\cite{Qutip1,Qutip2,Li2022} (see also the similar discussion in Refs.~\cite{Ku2020,Yang2022pra}).
In our noise model, we consider three different sources of the intrinsic noise from the device: qubit relaxation and qubit dephasing (QRQD), CNOT error, and readout error.

First, the QRQD is modeled by the following Lindblad master equation~\cite{Lindblad1976,Gorini1976}:
\begin{equation}
\begin{aligned}
\frac{\partial \rho(t)}{\partial t} &= \sum_{m}^{n}\frac{\gamma^{(m)}_{T_1}}{2}\left[2\sigma^{(m)}_-\rho(t)\sigma^{(m)}_+ - \{ \sigma^{(m)}_{-}\sigma^{(m)}_{+},\rho(t) \}\right] \\
&+\sum_{m}^{n}\frac{\gamma^{(m)}_{T_2}}{2}\left[2\sigma^{(m)}_z\rho(t)\sigma^{(m)}_z -\{\sigma^{(m)2}_z ,\rho(t)\}\right],
\end{aligned}\label{eq:Mastereq}
\end{equation}
where $\gamma^{(m)}_{T_1}=1/T^{(m)}_1$ and $\gamma^{(m)}_{T_2}=1/T^{(m)}_1-1/(2T^{(m)}_2)$ are the $m$th qubit relaxation and decoherence rates, respectively.
Here, $\sigma_{+} (\sigma_{-})$ denotes the atomic creation (annihilation) operator, and the corresponding relaxation (dephasing) time $T_1$ ($T_2$) are summarized in Table~\ref{Tab:IBM Q}.
We model the QRQD effect that occurs after performing a total unitary evolution (see \fig{\ref{fig:NoiseModel}}) and simulate it using the master equation solver MESOLVE in QuTip~\cite{Qutip1,Qutip2,Li2022}.
We sum over all the gate times in the circuit and obtain the total gate time $\approx3,725$~ns.
\begin{table}
    \begin{tabular}{ccccc}
        \hline\hline
        Qubits& $T_{1}$~($\upmu$s) &$T_2$~($\upmu$s)&$\Gamma_{X}~(10^{-4})$&$\Gamma_{\text{R}}~(10^{-2})$\\[3pt]  
        \hline
        B               & 118.4 & 194.5 & 1.5 & 1.0\\ [2pt]

        C               & 122.2 & 196.5 & 4.7 & 1.5\\ [2pt]

        $\text{E}_1$    & 84.1  & 44.1  & 1.7 & 0.1\\ [2pt]

        $\text{E}_2$    & 102.3  & 138.4 & 3.0 & 2.1\\ [2pt]
        \hline\hline\\[3pt]
        \hline\hline
         CNOT gate &\multicolumn{2}{c}{~~~CNOT error~$(10^{-3})$} &\multicolumn{2}{c}{Gate time (ns)}\\ [3pt]
        \hline
         $\text{B}~-~\text{C}$   &~~~~~~~~~~~~6.8  && ~~~~~~~~~ 309.3&\\ [2pt]
         ~$\text{B}~-~\text{E}_1$   &~~~~~~~~~~~~6.6  & & ~~~~~~~~~ 248.9&\\ [2pt]
         ~$\text{B}~-~\text{E}_2$   &~~~~~~~~~~~~9.0  & & ~~~~~~~~~ 202.7& \\ [2pt]
        \hline\hline
    \end{tabular}\caption{Error parameters in the IBM-Cairo device. The number of the four qubits in the IBM-Cairo device are labeled by: \#25, \#24, \#26, and \#22, representing the systems B, C, $\text{E}_1$, and $\text{E}_2$, respectively. Where $T_1$ ($T_2$) is the relaxation (dephasing) time, $\Gamma_X$ is the Pauli gate error, and $\Gamma_R$ is the readout error. Note that these error rates are public information on the IBM Q website~\cite{IBMWeb}. These numbers are presented here for completeness.\label{Tab:IBM Q}}
\end{table}
Note that each Pauli-$X$ gate in IBM-Cairo device takes $21.3$~ns, and the Hadamard gate $H$ (phase gate $S$) gate takes 5 (3) times longer than the Pauli-$X$ gate.

Second, the gate error is determined from the randomized benchmarking~\cite{PhysRevLett.106.180504,PhysRevA.85.042311}.
In a quantum assembly simulator, the gate error for the $n$-qubits system can be modeled by depolarizing noise~\cite{Urbanek2021prl}, i.e.,
\begin{equation}\label{eq:CNOTErr}
\begin{aligned}
\mathcal{G}_{\text{err}}(\rho)= (1-\Gamma_{\text{G}})\rho + \Gamma_{\text{G}} \frac{\id}{2^n},
\end{aligned}
\end{equation}
where $\Gamma_{\text{G}}$ is the gate error rate.
In our model, we assume that the gate errors are sequentially accumulated; thus, we multiply the different error rates which appear in the circuit.
Inserting the CNOT-gates error rate and the single-qubit Pauli-gates error rate $\Gamma_{X}$ shown in Table~\ref{Tab:IBM Q}, we estimate that this gate error rate is about $9.2\%$.

Finally, the readout error occurs because quantum devices have the probability of misrecording the ideal result 0(1) as 1(0).
Therefore, it can be modeled by a bit-flip channel, i.e., 
\begin{equation}\label{eq:ReadoutErr}
\begin{aligned}
\mathcal{R}_{\text{err}}(\rho) = (1-\Gamma_{\text{R}})\rho + \Gamma_{\text{R}}~\sigma_x \rho \sigma_x,
\end{aligned}
\end{equation}
where $\Gamma_{\text{R}}$ is the probability of the readout error.

As a result, the primary source causing errors is the number of CNOT gates, because they create a significant error rate compared to single-qubit gates. 
Moreover, the CNOT-gates also take longer time~\cite{Li2022}, meaning that they also increase the error effects from the QRQD.
Although we have ``only" used 12 CNOT-gates and 17 single-qubit gates in our circuit implementation of a superposition of the dephased phase shifts, it still creates errors greater than 27.0\%.

\section{Summary}\label{sec:conclusion}
In this paper, we generalize the metrological steering task described in Ref.~\cite{NC2021:BYadin} to a scenario with superpositions of noisy phase shifts.
We show that the control in $\ket{+}\bra{+}_{\text{C}}$ (i.e., via a superposition of dephased and depolarized phase shifts) can alleviate the noisy effect and enhance the average violations of the MSI in comparison with the case where the control is in an incoherent mixed state (i.e., without superposition of dephased and depolarized phase shifts).

Moreover, we proposed a circuit model for superposing two dephased phase shifts and experimentally implemented the circuit on the IBM Quantum Experience.
We clearly observe the violations of the MSI, and the experimental results agree with our noise simulations.

Finally, it is known that the order of channels can also be coherently controlled~\cite{PRL2018:EDaniel,Barrett2021}.
Therefore, it would be promising to apply this framework to the noisy metrological steering task.

\section*{Acknowledgments}
We acknowledge fruitful discussions with Yi-Te Huang and Feng-Jui Chan.
We acknowledge the NTU-IBM Q Hub and the IBM Quantum Experience for providing us a platform to implement the experiment.
The views expressed are those of the authors and do not reflect the official policy or position of IBM or the IBM Quantum Experience team.
A.M. is supported by the Polish National Science Centre (NCN) under the Maestro Grant No. DEC-2019/34/A/ST2/00081.
F.N. is supported in part by: Nippon Telegraph and Telephone Corporation (NTT) Research, the Japan Science and Technology Agency (JST) [via the Quantum Leap Flagship Program (Q-LEAP), and the Moonshot R\&D Grant Number JPMJMS2061], the Japan Society for the Promotion of Science (JSPS) [via the Grants-in-Aid for Scientific Research (KAKENHI) Grant No. JP20H00134], the Army Research Office (ARO) (Grant No. W911NF-18-1-0358), the Asian Office of Aerospace Research and Development (AOARD) (via Grant No. FA2386-20-1-4069), and the Foundational Questions Institute Fund (FQXi) via Grant No. FQXi-IAF19-06.
H.-Y. K. is supported by the National Center for Theoretical Sciences and National Science and Technology Council, Taiwan (Grants MOST No. 110-2811-M-006-546 and MOST No. 111-2917-I-564-005).
This work is supported by the National Center for Theoretical Sciences and National Science and Technology Council, Taiwan, Grants No. MOST 111-2123-M-006-001.

\appendix
\section{Circuit model of a superposition of depolarized phase shifts}\label{sec:appendix}
\begin{figure*}[!htbp]
    \includegraphics[width=1.99\columnwidth]{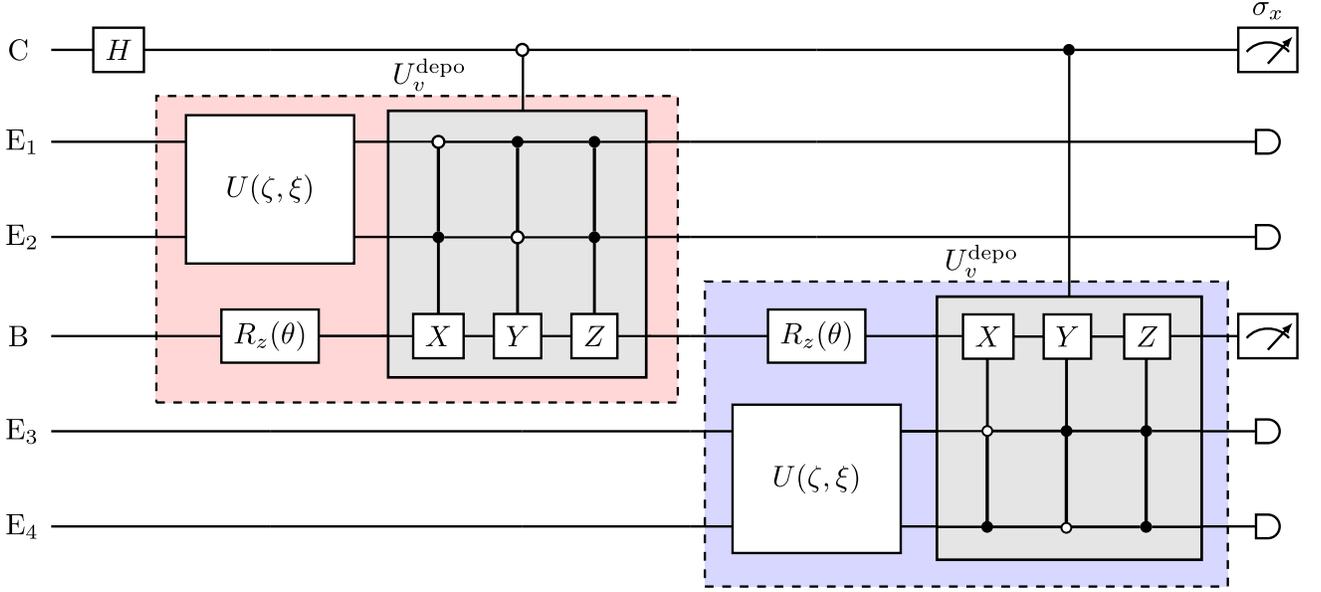}
    \caption{Circuit implementation of a superposition of depolarized phase shifts. We assume that the control system C and all the environments $\text{E}_i$ $(i=1, ..., 4)$ are initialized to $\ket{0}$.}\label{dex:Depcirc}
\end{figure*}
In this Appendix, we aim to construct a circuit that satisfies the depolarized phase shifts implementing the operations in Eq.~\eqref{Sim:Depolarizing Unitary}.
A direct way to design a depolarizing phase shift circuit is that we can use three different kinds of Toffoli gates to represent the system transformation errors modeled by $\sigma_x$, $\sigma_y$, and $\sigma_z$ with different probabilities~\cite{QIQC}.
As shown in Fig.~\ref{dex:Depcirc}, we use a two-qubit system, which plays the role of a four-level environment in Eq.~\eqref{Sim:Depolarizing Unitary}, i.e.,
\begin{equation}
    \begin{aligned}
        \ket{0}_{\text{E}}&\rightarrow{}\ket{0}\ket{0}_{\text{E}},\quad
        \ket{1}_{\text{E}}\rightarrow{}\ket{0}\ket{1}_{\text{E}},\\
        \ket{2}_{\text{E}}&\rightarrow{}\ket{1}\ket{0}_{\text{E}},\quad
        \ket{3}_{\text{E}}\rightarrow{}\ket{1}\ket{1}_{\text{E}}.
    \end{aligned}
\end{equation}

\begin{figure}[!htbp]
    \includegraphics[width=0.95\columnwidth]{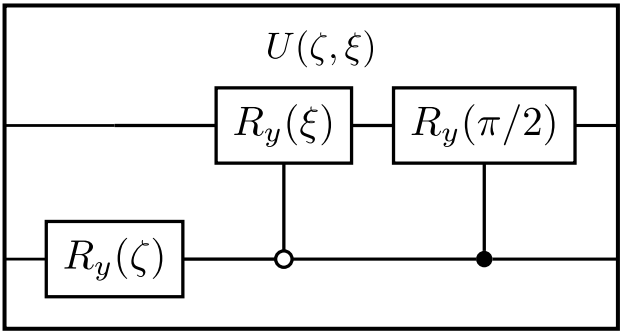}
    \caption{Circuit implementing the unitary operation $U(\zeta,\xi)$, which maps $\ket{0}\ket{0}_{\text{E}}$ onto the state in Eq.~\eqref{app:dep state}.\label{dex:DepPrep}}
\end{figure}

To fit the factors $\sqrt{1- 3v/4}$ and $\sqrt{v/4}$ in Eq.~\eqref{Sim:Depolarizing Unitary}, we apply a unitary $U(\zeta,\xi)$, which maps the two-qubit environment $\ket{0}\ket{0}_{\text{E}}$ into
\begin{equation}
    \sqrt{1- \frac{3v}{4}}\ket{0}\ket{0}_{\text{E}}+\sqrt{ \frac{v}{4}}\left(\ket{0}\ket{1}_{\text{E}}+\ket{1}\ket{0}_{\text{E}}+\ket{1}\ket{1}_{\text{E}}\right)
\end{equation}
with two rotation parameters $\zeta$ and $\xi$ on the environmental system (see also Fig.~\ref{dex:DepPrep}).
After mapping $U(\zeta,\xi)$, we obtain the initial state  
\begin{equation}
\begin{aligned}
    \ket{0}\ket{0}_{\text{E}} \rightarrow{}  &\cos{\frac{\zeta}{2}}\cos{\frac{\xi}{2}} \ket{0}\ket{0}_{\text{E}} +\sqrt{\frac{1}{2}}\sin{\frac{\zeta}{2}} \ket{0}\ket{1}_{\text{E}}\\ +&\cos{\frac{\zeta}{2}}\sin{\frac{\xi}{2}} \ket{1}\ket{0}_{\text{E}} +\sqrt{\frac{1}{2}}\sin{\frac{\zeta}{2}} \ket{1}\ket{1}_{\text{E}}.\label{app:dep state}
\end{aligned}
\end{equation}
One can find that if we let $\zeta=2\text{sin}^{-1}\sqrt{v/2}$ and $\xi=2\text{sin}^{-1}\sqrt{v/(4-2v)}$, we can obtain the red (blue) box in \fig{\ref{dex:Depcirc}}, which is equal to $U^{\text{depo}}_{v}$ in Eq.~\eqref{Sim:Depolarizing Unitary}.

In general, to implement a superposition of quantum channels in a gate-based quantum simulation requires using many Toffoli gates~\cite{MICHIELSEN201744,Fedorov_2011}.
For the superposition of two depolarized phase shifts, we require an additional control system. Therefore, there are six controlled Toffoli gates required to simulate the desired dynamics (see Fig.~\ref{dex:Depcirc}).
Since a Toffoli gate can be decomposed into six CNOT gates and nine single-qubit gates~\cite{QIQC}; therefore, a single controlled Toffoli gate contains 52 CNOT gates and needs $\approx 16,400$~ns to operate.

In total, there are 328 CNOT gates in our circuit of depolarized phased shifts, creating gate-error rates of at least 94.3\%, and a total gate time $\approx 111,945$~ns.
In our noise simulations of the depolarized noise phase shifts, the $4\tilde{\Delta}Z^{\text{avg}}_{\text{opt}}$ is larger than 0.99 and the $F^{\text{avg}}_{\text{opt}}$ is less than 0.01.
Thus, we do not observe the violation of the metrological steering inequality in Eq.~\eqref{eq:Steering ineq} on IBM Q devices because the circuits error is too large and destroys the quantum advantages. 

For clarity and completeness, we recall the meaning of standard gates used in our implementation both in Figs.~\ref{fig:IBM_Expsetting} and \ref{dex:Depcirc}.
Specifically, $X$, $Y$, $Z$ represent Pauli gates, $H$ the Hadamard gate, and $S$ the phase gate, defined as $S={\rm diag}(1,i)$.
Also, $R_z(\theta)$ is the rotation gate along the $z$-axis with angle $\theta$, written as $R_z(\theta)={\rm diag}[\exp{(-i\theta/2)},\exp{(i\theta/2)}]$.



%


\end{document}